\documentclass[12pt]{iopart}
\usepackage{iopams}  
\usepackage{graphicx}
\begin{document}

\newcommand{\HFBTHO}{\textsc{hfbtho}}
\newcommand{\UNEDFZERO}{\textsc{unedf0}}
\newcommand{\UNEDFONE}{\textsc{unedf1}}
\newcommand{\UNEDFTWO}{\textsc{unedf2}}
\newcommand{\UNEDF}{\textsc{unedf}}

\newcommand{\asym}{a_{\rm sym}^{\rm NM}}
\newcommand{\lsym}{L_{\rm sym}^{\rm NM}}
\newcommand{\nskin}{$r_{\rm ns}$}

\newcommand{\algo}{\textsc{pound}er\textsc{s}}

\title{Propagation of uncertainties in the nuclear DFT models}

\author{Markus Kortelainen}
\address{Department of Physics, P.O. Box 35 (YFL), University of Jyv\"askyl\"a, FI-40014 Jyv\"askyl\"a, Finland}
\address{Helsinki Institute of Physics, P.O. Box 64, FI-00014 University of Helsinki, Finland}
\ead{markus.kortelainen@jyu.fi}

\begin{abstract}
Parameters of the nuclear density functional theory (DFT) models are usually adjusted to experimental
data. As a result they carry certain theoretical error, which, as a consequence, carries out to the
predicted quantities. In this work we address the propagation of theoretical error, 
within the nuclear DFT models, from the model parameters to the predicted observables.
In particularly, the focus is set on the Skyrme energy density functional models.
\end{abstract}

\pacs{21.10.-k, 21.30.Fe, 21.60.Jz}

\maketitle

%
%
\section{Introduction}

The purpose of the theoretical models is not only to explain current observations, but also to make
predictions to be tested with the future observations. In nuclear physics particularly, there still exists a wide
gap between experimentally know nuclei and the expected limits of the nuclear landscape, in particularly for
the heavier, very neutron rich nuclei \cite{(Erl12)}. Even though these nuclei are yet to be observed
in the laboratory, many of them have impact on astrophysical scenarios \cite{(Apr14)}.
To predict properties of nuclei, various theoretical models have been developed and applied.
In extrapolation to experimentally unknown region, the role of model uncertainties becomes prominent.
Without accompanied error analysis, the model prediction bears no meaning.

Presently, the nuclear density functional theory (DFT) is the only microscopic theory which can
be applied throughout the entire nuclear chart. The key ingredient of the DFT is the energy density
functional (EDF), which incorporates complex many-body correlations into a functional, which is
constructed from the nucleonic densities and currents. 
Currently, there are three main variants of the nuclear DFT; zero-range Skyrme, finite range Gogny,
and relativistic mean field models \cite{(Ben03)}. In this work we focus on the Skyrme-EDF models.
Similar kind of analysis for other nuclear DFT variants could be also done.

Similarly like with any other effective theory, the parameters of the nuclear DFT models need to be 
adjusted to empirical input. 
Historically, in the optimization of various nuclear DFT models, covariance error analysis of the obtained
parameterization was usually neglected. However, as stressed for example in \cite{(Toi08),(Dob14)}, error
analysis is an essential post-optimization tool to quantify the model errors.
Only very recently, such kind of analysis has been performed for nuclear DFT models
\cite{(Klu09),(Kor10),(Fat11),(Kor12),(Erl13),(Kor14)}. This kind of information is vital
when assessing predictive power of the model.

Theoretical uncertainties related to the limits of the nuclear landscape, within Skyrme-EDF models,
were studied in \cite{(Erl12)}. It was found that, despite large differences in various
optimization procedures, current Skyrme-EDF models give a rather consistent picture about the
position of the proton and neutron drip-lines. Furthermore, it was found that both, systematic model
error and statistical model error, gave rather similar uncertainty width for the drip-line.
Similar studies for relativistic mean-field models were also done in \cite{(Afa13),(Agb14)}. 
The results were rather similar to Skyrme-EDF results.
Uncertainties within the Brussels-Montreal HFB mass models were analyzed in \cite{(Gor14)}.

With a given covariance matrix of the model, the propagation of the error from the model 
parameters to the model predictions can be computed. 
The information content and effect of a new observable to the uncertainties of the model 
parameters and predictions was investigated in \cite{(Rei10)}. It was found that a precise
measurement of neutron skin thickness in $^{208}$Pb could reduce model errors of other isovector observables.
In \cite{(Gao13)} propagation of the error within {\UNEDFZERO} EDF \cite{(Kor10)} model was 
investigated in semi-magic nuclei. Statistical and systematical model errors 
related to the neutron skin thickness were studied in \cite{(Kor13)}. In both of these studies,
the poorly constrained isovector parameters were found to be the largest source to the model error.

This article is organized as follows: In Section \ref{Sec:Theory} we review theoretical
formalism. Some examples are discussed in Section \ref{Sec:Results}, and in Section
\ref{Sec:Conclusions} we present concluding remarks.

%
%
\section{Theoretical framework} \label{Sec:Theory}
\subsection{Skyrme energy density}
In the Skyrme-EDF framework the total binding energy $E$ of the nucleus is
\begin{equation}
E = \int d^{3}r \left( {\mathcal E}^{\rm Kin}({\bi r}) + {\mathcal E}^{\rm Sk}_{0}({\bi r}) + {\mathcal E}^{\rm Sk}_{1}({\bi r}) 
+{\mathcal E}^{\rm Pair}({\bi r}) + {\mathcal E}^{\rm Coul}({\bi r}) \right) \, ,
\end{equation}
where ${\mathcal E}^{\rm Kin}({\bi r})$ is the kinetic energy density, ${\mathcal E}^{\rm Pair}({\bi r})$
is the pairing energy density, and ${\mathcal E}^{\rm Coul}({\bi r})$ is the Coulomb energy density.
Isoscalar ($t=0$) and isovector ($t=1$) time-even part of the Skyrme energy density is defined as
\begin{equation} \label{eq:Skyrme}
{\mathcal E}^{\rm Sk}_{t}({\bi r}) = C^{\rho}_{t}[\rho] \rho_ {t}^{2} + C^{\tau}_{t} \rho_{t}\tau_{t} 
+C^{\Delta\rho}_{t} \rho_{t} \Delta \rho_{t} + C^{\nabla J}_{t} \rho_{t} \nabla \cdot {\bi J}_{t}
+C^{J}_{t} {\mathbb J}^{2}_{t} \, ,
\end{equation}
where density dependency is 
\begin{equation}
C^{\rho}_{t}[\rho] = C^{\rho}_{t0}+C^{\rho}_{t{\rm D}}\rho^{\gamma}_{0} \, .
\end{equation}
In equations above, we have omitted, for brevity, dependence of densities on the spatial coordinate ${\bi r}$. 
The Skyrme energy density (\ref{eq:Skyrme}) is composed of matter density $\rho_{t}$, kinetic density $\tau_{t}$,
spin-current vector density ${\bi J}_{t}$, and spin-current tensor density ${\mathbb J}_{t}$. Isoscalar
matter density is defined as $\rho_{0}=\rho_{\rm n}+\rho_{\rm p}$, and isovector matter density as
$\rho_{1}=\rho_{\rm n}-\rho_{\rm p}$, where ${\rm n}$ and ${\rm p}$ are neutron and proton indices, respectively.
Isoscalar and isovector density is defined similarly also for the kinetic and spin-current densities.
Standard definitions of these densities can be found e.g. in \cite{(Ben03),(Dob96b)}.
Here, $C^{\rho}_{t}[\rho] \rho_ {t}^{2}$ is the local volume term, which gives the largest contribution coming from 
the energy density to the total binding energy. Term $C^{\tau}_{t} \rho_{t}\tau_{t}$ is connected to the effective mass.
Term $C^{\Delta\rho}_{t} \rho_{t} \Delta \rho_{t}$ is the local surface term, which mostly contributes at the surface
of the nucleus.
Furthermore, $C^{\nabla J}_{t} \rho_{t} \nabla \cdot {\bi J}_{t}$ and $C^{J}_{t} {\mathbb J}^{2}_{t}$ are
the spin-orbit and tensor terms, respectively. The effect of spin-orbit term (and tensor term to lesser extend) 
is to push spin-orbit partner orbitals apart, to reproduce correct magic numbers.

The Skyrme energy density of equation (\ref{eq:Skyrme}) is parameterized by the coupling constants $C^{i}_{t}$.
At the present, these parameters can not be pre-calculated from any theory within sufficient accuracy and, 
therefore, they need to be adjusted to empirical input.
There is also a one-to-one correspondence between Skyrme coupling constants of the volume part,
that is $\left\{ C^{\rho}_{t0},C^{\rho}_{t{\rm D}}, C^{\tau}_{t}, \gamma; \, t=0,1 \right\}$, 
and infinite nuclear matter (INM) parameters~\cite{(Agr05),(Kor10)}. 
Corresponding INM parameters have been listed on Table \ref{tt:INM}.
INM parameters present a more convenient way to parameterize a part of the EDF since their values are approximately known. 

\begin{table}
\caption{\label{tt:INM}Infinite nuclear matter parameters.}
\begin{indented}
\item[]\begin{tabular}{@{}ll}
\br
Symbol & Explanation \\
\mr
$\rho_{\rm c}$ & Equilibrium density of the nuclear-matter \\ $E^{\rm NM}/A$ & Total energy per nucleon at equilibrium \\
$K^{\rm NM}$ & Nuclear-matter incompressibility  \\ $m^{*}_{\rm s}/m$ & Scalar effective mass \\
$\asym$ & Symmetry energy coefficient \\ $\lsym$ & Density dependence of the symmetry energy \\
$m^{*}_{\rm v}/m$ & Vector effective mass \\
\br
\end{tabular}
\end{indented}
\end{table}

To solve the ground state wave-function, within the single-reference Skyrme-EDF framework, one has to solve 
the Hartree-Fock or Hartree-Fock-Bogoliubov equations self-consistently. This procedure is outlined, e.g., 
in \cite{(Ben03),(Rin00)}.

\subsection{Optimization of the model parameters}
With the DFT models used in nuclear theory, the
model parameters are usually adjusted to empirical input, and in some occasions, to some other pseudo-data not directly 
related to experimental result.
Typical adjustment involves a minimization of the $\chi^{2}$ function, which is usually defined as
\begin{equation} \label{eq:chi2}
\chi^{2}({\bi x}) = \sum_{i=1}^{n_{\rm d}} \left( \frac{s_{i}({\bi x}) - d_{i}}{w_{i}} \right)^2 \, ,
\end{equation}
where ${\bi x}$ is an array of model parameters, $s_{i}({\bi x})$ is the value of data point calculated from the model,
and $d_{i}$ is the corresponding empirical value for data point. Furthermore, $n_{\rm d}$ is the number of data
points, and $w_{i}$ is the used weight of the data point. The definition
$\chi^{2}({\bi x})$ is such that it is a unitless function. To obtain the minimum of the $\chi^{2}({\bi x})$,
at ${\bi x}={\bi x}_{\rm min}$, various algorithms can be used. In optimization of {\UNEDF} energy density
functionals {\algo} algorithm was used, which offers significant improvement in terms of used CPU time compared to
traditional Nelder-Mead algorithm \cite{(Kor10)}.

As for the empirical input, nuclear DFT models have been typically optimized to various ground-state properties,
see e.g. review in \cite{(Ben03)}. In optimization of {\UNEDFZERO} \cite{(Kor10)}, binding energies, 
charge radii, and odd-even staggering data was used. 
For {\UNEDFONE} \cite{(Kor12)} optimization, data on the fission isomer excitation energies were added to the 
pool of data points. As a result, fission properties were significantly improved. With {\UNEDFTWO} \cite{(Kor14)}, 
shell structure was addressed by including data on single-particle level energies.

\subsection{Determination of covariance matrix}
The calculated model predictions $s_{i}({\bi x})$ of equation (\ref{eq:chi2}) are, in nuclear DFT models, usually non-linear functions
with respect of model parameters ${\bi x}$. In this kind of situation, a rigorous calculation of the covariance matrix is a
formidable task, yet to be done. Therefore, usually a linearized least square system in the vicinity 
of the minimum ${\bi x}_{ \rm min}$ of the object $\chi^{2}({\bi x})$ function is assumed. Within this approximation, the
$s_{i}({\bi x})$ is assumed to be a linear function of ${\bi x}$. The validity of this approximation depends
on the type of the observable and investigated width of the landscape around the minimum.
For example, single-particle energies behave rather linearly with respect of Skyrme coupling constants \cite{(Kor08),(Tar14)}.
Also, since the total energy $E$ in the Skyrme-EDF scheme is directly proportional to the coupling constants, as shown in 
equation (\ref{eq:Skyrme}), it should also behave relatively linearly. 
This apparent linearity was also used in \cite{(Ber05),(Sto10)} for parameter optimization.
Nevertheless, non-linearities appear in self-consistent picture, where calculated densities depend on the coupling constants.
A linearized approximation to covariance matrix is also numerically more stable compared to alternative choices \cite{(Kor10)}.

Within aforementioned linearized approximation, covariance matrix of the model can be written as a matrix equation \cite{(Dob14),(Bra99)}
\begin{equation} \label{eq:cov}
{\rm Cov}(x_{i},x_{j})  =  \left[ \left(A^{T} G_{w} A\right)^{-1} \right]_{ij}\, , 
\end{equation}
where
\begin{eqnarray}
A_{ij}  &=&  \left.\frac{\partial s_{i}({\bi x})}{\partial x_{j}}\right|_{{\bi x}={\bi x}_{\rm min}} \, , \label{eq:Aij}\\
G_{w}   &=&  {\rm diag}(w_{1}^{-2},\ldots,w_{n_{d}}^{-2}) \, ,
\end{eqnarray}
where the weights $w_{i}$ are the same weights as used in the $\chi^{2}({\bi x})$ of equation (\ref{eq:chi2}). Derivatives of equation (\ref{eq:Aij}) 
typically need to be calculated numerically e.g. by taking a finite difference. However, in some cases a closed analytic
form may be available.
The normalization of $\chi^{2}({\bi x})$ of equation (\ref{eq:chi2}) and covariance matrix (\ref{eq:cov}) is selected according
to ''simple'' convention of \cite{(Dob14)}. See the Appendix of \cite{(Dob14)} for other conventions.

Closely connected to the covariance matrix is the correlation coefficient between two model parameters, which is defined as
\begin{equation}
R_{ij} = \frac{{\rm Cov}(x_{i},x_{j})}{\sqrt{{\rm Var}(x_{i}){\rm Var}(x_{j})}} \, ,
\end{equation}
where ${\rm Var}(x_{i})=\sigma_{i}^{2}$, and $\sigma_{i}$ is the standard deviation
of model parameter $x_{i}$. A strong correlation coefficient between two model parameters indicates that
the covariance ellipsoid of the $\chi^{2}({\bi x})$ landscape, close to the minimum, is strongly elongated \cite{(Rei10)}.

\begin{figure}[htb]
\center
\includegraphics[width=0.66\linewidth]{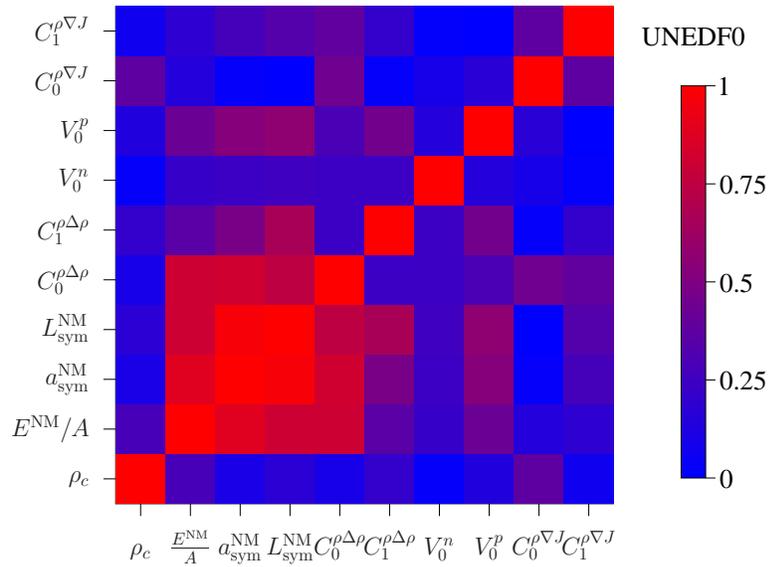}
\caption{Correlation matrix of {\UNEDFZERO} parameterization presented graphically.
Shown are absolute values of correlation coefficients.}
\label{fig:unedf0correl}
\end{figure}

As an example, figure~\ref{fig:unedf0correl} shows the correlation matrix of {\UNEDFZERO} parameterization~\cite{(Kor10)}.
From the figure, one can see that some of the volume parameters are strongly correlated with each other, as well as
with the isoscalar surface coupling constant $C^{\rho\Delta\rho}_{0}$.

\subsection{Propagation of error}
With the covariance matrix known, the standard deviation $\sigma(O)$ of an observable $O$ can be calculated as
\begin{equation} \label{eq:stddev}
\sigma^{2}(O) = \sum_{i,j} {\rm Cov}(x_{i},x_{j})\left.\left(\frac{\partial O(\bi{x})}{\partial x_{i}}\right)
\left(\frac{\partial O(\bi{x})}{\partial x_{j}}\right) 
\right|_{ {\bi x}={\bi x}_{\rm min} }\, ,
\end{equation}
where $O(\bi{x})$ is the predicted value of the observable  by the model. Again, as with
the calculation of the covariance matrix, derivatives with respect to model parameters 
usually need to be  calculated numerically, by taking a finite difference. 
The standard deviation $\sigma(O)$ can be also referred, loosely speaking, as a "theoretical error".
However, one should bear in mind, that there are also many other sources of theoretical errors,
and it may be very difficult task to estimate the total combined theoretical error.

\subsection{Selection of weights} \label{Sec:TheoryWeights}
Up to so far, we have not defined what the weights $w_{i}$ of equation (\ref{eq:chi2}) should be.
The selection of the weights brings some arbitrariness to the adjustment of the model parameters.
The position of the minimum ${\bi x}_{\rm min}$ depends on the relative magnitude of the used weights
between various data points. By changing relative weighting between, e.g., binding energies and
charge radii, the EDF model could be tailored to reproduce this kind of data more accurately.
As a consequence, in the various EDF optimization schemes, there has not been a common practice regarding
of selection of the weights.

To reduce this kind of arbitrariness,
one option would be to use experimental error bars as a weights of the data points in equation (\ref{eq:chi2}).
However, in the light of the performance of current EDF models, this is hardly a viable solution.
Experimental mass measurements with Penning-traps can reach up to sub-keV level in accuracy~\cite{(Kan12)}, which
is orders of magnitude more precise compared to expected accuracy of roughly $1\,{\rm MeV}$ of the present day mean-field models.
With single particle (s.p.) energies, the situation is the same. Current Skyrme-like functionals can reproduce 
empirical s.p. data only with a typical root-mean-square (r.m.s.) deviation of roughly $1\,{\rm MeV}$~\cite{(Kor08),(Kor14),(Tar14)}.

The other alternative is to tune the weights according to the capabilities of the used theoretical framework.
As argued in \cite{(Gao13)}, the weights should reflect expected accuracy of the model.
Indeed, this has been the usual choice in many recent Skyrme-EDF optimization schemes. For example,
in optimization of {\UNEDFZERO}, the weights for nuclear binding energy data points were set
to $2\,{\rm MeV}$, and for the charge radii $0.02\,{\rm fm}$. This corresponds rather well the
overall performance on {\UNEDFZERO}: The r.m.s. deviation for binding energies, 
over the whole even-even mass table, was $1.4\,{\rm MeV}$ and for the charge radii $0.017\,{\rm fm}$.
At the present, due to the lack of any better guideline, it seems that the most appropriate way to set the
weights is to use expected accuracy of the model for a each given data type.

\section{Some examples} \label{Sec:Results}

In this section, some illustrative examples about the propagation of theoretical error are shown.
Examples are mainly calculated with {\UNEDFZERO} Skyrme-EDF. However, similar studies can
be also done with other EDFs where the full covariance matrix of the model parameters is known.

\subsection{Binding energies and separation energies}

One of the most rudimentary observable, which a universal nuclear EDF is expected to reproduce
well across the nuclear chart, is the binding energy. As discussed in Sec.~\ref{Sec:TheoryWeights},
presently nuclear DFT models have a typical r.m.s. deviation from experimental data of the order of
$1\,{\rm MeV}$. Since the calculated total energy $E$, within the Skyrme-EDF model, is directly connected
to the model parameters, the uncertainty of these parameters propagates to the predicted energy.

\begin{figure}[tbh]
\center
\includegraphics[width=0.66\linewidth]{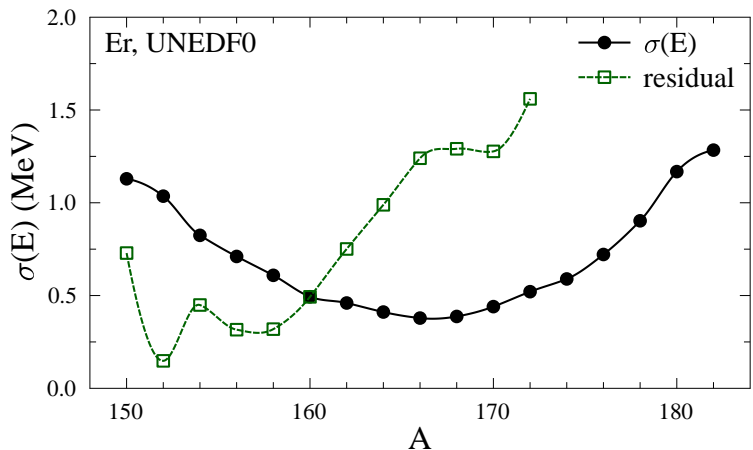}
\caption{Calculated theoretical standard deviation $\sigma(E)$ of binding energies 
together with the absolute values of residuals to the experimental data in Er isotopic chain,
with {\UNEDFZERO} EDF. All in units of MeV.}
\label{fig:beEr}
\end{figure}

Figure~\ref{fig:beEr} shows the calculated theoretical error in predicted binding energies of even-even
erbium isotopes within {\UNEDFZERO} EDF model. Shown are also absolute values of residuals compared to 
experimental data of \cite{(Aud03)}, that is, the difference between experimental and theoretical result.
The results, as well as the results of Sec.~\ref{Sec:ResultsNSkin}, were calculated with computer code
{\HFBTHO} \cite{(Sto13)}.
Since the volume part of {\UNEDFZERO} is parameterized with INM parameters, as well as corresponding part of the covariance matrix, 
they are also used here in calculation of derivatives of equation (\ref{eq:stddev}). Tensor coupling
constants $C^{J}_{t}$ in the {\UNEDFZERO} EDF model are set to zero.
Propagation of error to the binding energies of semi-magic
isotopic and isotonic chains was also studies in \cite{(Gao13)}. Similarly as in 
\cite{(Gao13)}, the propagated error here increases when moving further out from the
valley of stability. Relation of the model residuals to experimental data was also investigated
for lead isotopic chain, and it was found that around double-magic $^{208}{\rm Pb}$, the residuals
were much larger compared to propagated error. This, and the fact that residuals along the isotopic 
chains are nor randomly distributed, but usually show arc-like features (see e.g. \cite{(Kor10)}),
are clear examples of the deficiency of the model.
Here, in Er isotopic chain, the difference between residuals and propagated error is not as drastic, 
due to the fact that this kind of Skyrme EDF models are expected perform better in deformed open 
shell nuclei.

\begin{figure}[tbh]
\center
\includegraphics[width=0.66\linewidth]{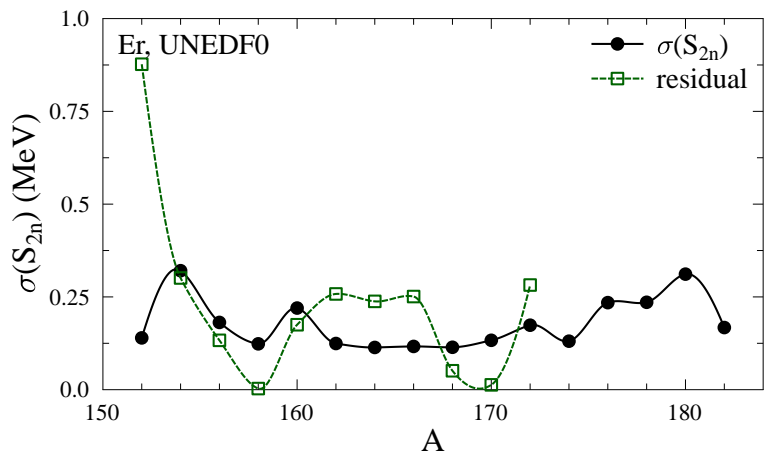}
\caption{The same as figure~\ref{fig:beEr}, but for the two-neutron separation energies.}
\label{fig:s2nEr}
\end{figure}

Theoretical error for two-neutrons separation energies is shown in figure~\ref{fig:s2nEr} along
with the residuals to experimental values for {\UNEDFZERO}. As can be seen, the magnitude
of residuals and the magnitude of theoretical error is usually rather similar. This is an indication
that the the weights selected for {\UNEDFZERO} optimization are in balance with expected accuracy
of the model. The only notable deviation exists at the magic neutron number $N=82$.  
Theoretical error for two-neutron separation energy was also studied in \cite{(Erl12)} 
for SV-min~\cite{(Klu09)} EDF. Similarly as with SV-min, the error increases towards neutron rich 
regime. The same was also found in \cite{(Gao13)}.

\subsection{Neutron skin thickness}\label{Sec:ResultsNSkin}
Recently, there has been a considerable theoretical and experimental interest on the neutron
skin thickness. 
Neutron skin is defined as a neutron matter distribution extending further out compared to proton matter distribution.
Neutron skin can be characterized by its thickness, which correlates strongly with other isovector
observables \cite{(Bro00),(Fur02),(Rei10),(Roc11),(Pie12),(Rei13),(Rei13b),(Naz14)}. It has also a strong connection to neutron
matter equation of the state \cite{(War09),(Fat12),(Agr12),(Lat12),(Ste13),(Erl13)}.
Experimentally, various probes have been used to determine neutron skin thickness. 
A recent measurement, the Lead Radius Experiment (PREX) \cite{(Abr12)}, determined the parity-violating asymmetry coefficient 
in $^{208}$Pb, which is directly related to the neutron skin thickness providing $r_{\rm ns}=0.33^{+0.16}_{-0.18}\,{\rm fm}$ \cite{(Hor12)}.
This is probably the most model independent measurement of the neutron skin thickness.

Error propagation to predicted neutron skin thickness was studied in \cite{(Kor13)}, where both, systematic and statistical,
model error were addressed. One of the major outcomes of \cite{(Kor13)} was that the statistical model
error is the defining theoretical uncertainty in the case of neutron skin thickness, whereas systematic model
error was found to be notably lower.

\begin{figure}[tbh]
\center
\includegraphics[width=1.0\linewidth]{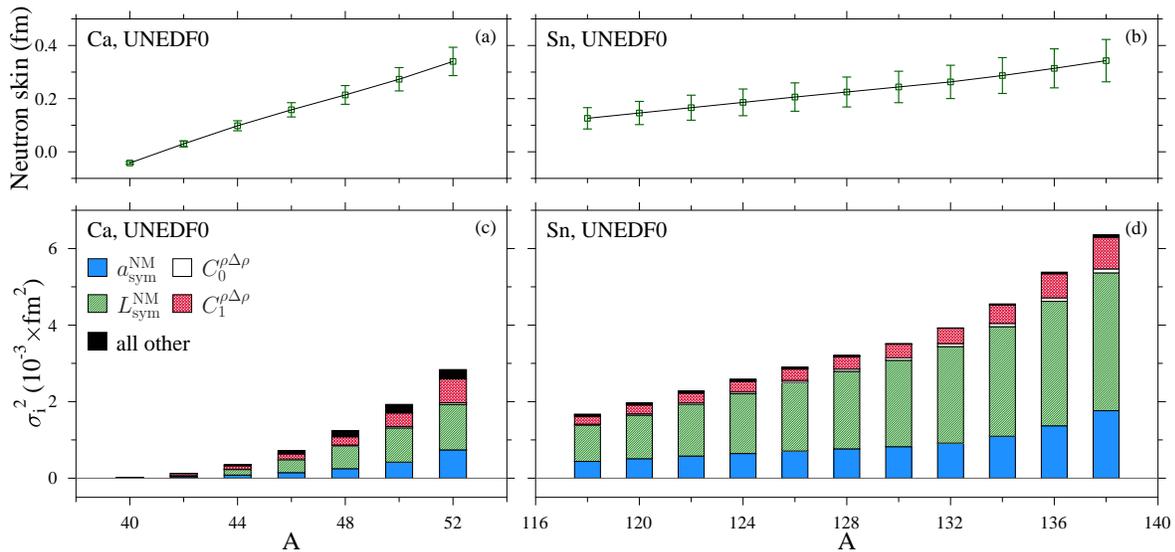}
\caption{Upper panels: Predicted neutron skin thickness with theoretical uncertainty
for Ca (a) and Sn (b) isotopic chains in units of fm. Lower panels: Theoretical error
budget for neutron skin uncertainty for Ca (c) and Sn (d) isotopic chains. All calculated 
with {\UNEDFZERO} EDF.}
\label{fig:nskin-ca-sn}
\end{figure}

Figure~\ref{fig:nskin-ca-sn} shows predicted neutron skin thickness in even-even calcium and tin
isotopes together with theoretical statistical error and error budget.
Both, neutron skin thickness and related theoretical uncertainty, increase towards more neutron
neutron rich isotopes.
As shown on the lower panels, the density dependence of the symmetry energy, $\lsym$, has the
largest contribution to the total error of the neutron skin thickness in both, calcium and tin
isotopes. The situation was found to be the same also for lead isotopes in \cite{(Kor13)}.
The second largest contributor comes from symmetry energy, $\asym$, similarly again as was found in \cite{(Kor13)}.
This is one example, among many others, about the poorly determined isovector parameters in the EDF models.

\begin{figure}[tbh]
\center
\includegraphics[width=0.33\linewidth]{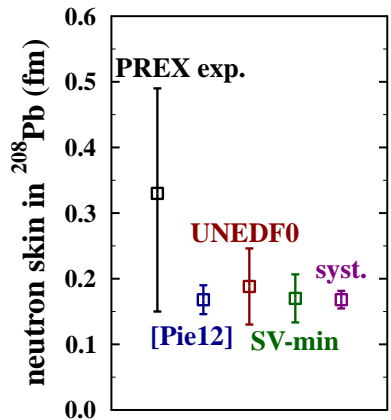}
\caption{Experimental PREX result and theoretical model predictions with error bars for the neutron skin thickness in $^{208}{\rm Pb}$. 
Label [Pie12] refers to systematic model average of \cite{(Pie12)}. {\UNEDFZERO}, SV-min and systematic model error 
are taken from \cite{(Kor13)}. All in units of fm.}
\label{fig:nskin208pb}
\end{figure}

Current status related to the uncertainties of the neutron skin thickness in $^{208}{\rm Pb}$ is summarized in
figure~\ref{fig:nskin208pb}, which shows the result of PREX experiment together with systematic model error of
\cite{(Pie12)}, and the systematic and statistical model error obtained in \cite{(Kor13)}. The statistical
model error is shown for {\UNEDFZERO} and SV-min EDFs. As discussed in \cite{(Kor13)} 
the PREX measurement, with a rather large error bar, can not constrain current EDF models. Planned
PREX-II \cite{Prex2} and CREX \cite{crex} experiments, which will measure neutron skin thickness in $^{208}{\rm Pb}$ and $^{48}$Ca, 
aim at much higher experimental accuracy. Results provided by these experiments could help to constrain isovector 
parameters in the future EDF optimization schemes.

\section{Concluding remarks} \label{Sec:Conclusions}

In this article, we have analyzed theoretical uncertainties related to the nuclear EDF models.
In particularly, the focus has been on the propagation of the error to the predicted observables
within the framework of {\UNEDFZERO} Skyrme EDF. Nuclear binding energies, two-neutron separation
energies and neutron skin thickness were used as an examples.
In all cases, as well as in the previous studies, it was found that theoretical error increases
towards increasing isospin excess. This has been directly linked to the poorly known isovector parameters of
the EDFs.
Similar kind of analysis could be done for other EDF models, where information about the covariance matrix exists.
Nevertheless, with this kind of analysis, one should always bear in mind, that the procedure outlined in this
article can only assess theoretical uncertainties within the model itself. Uncertainties and model errors
connected to the deficiency of the model itself can not be estimated with this procedure. One signature of such kind of deficiency
is a clear pattern of residuals of experimental data, which are significantly larger compared to calculated
theoretical model error. In \cite{(Gao13)} binding energies close to double-magic $^{208}{\rm Pb}$,
and single-particle energies were found to follow this kind of pattern. Yet another source of errors
are the used numerical and computational procedures.  

In the future EDF parameter optimization schemes, sensitivity analysis of the obtained parameterization
will be prime importance. Without such a procedure, there is no handle to judge the predictive 
power of the model. As concluded in \cite{(Kor14)}, the limits of the Skyrme-like EDFs have been 
reached and major improvements are not to be expected on this path.
To go beyond present day EDF models, many alternatives have been presented \cite{(Car08),(Sto10),(Rai14)}. 
Nonetheless, all of them share the common feature of the requirement to adjust the model parameters to 
empirical input. It will be exciting to see how these new EDF models perform across the nuclear chart
and how good will be their predictive power.

\section*{Acknowledgments}
This work was supported by the Academy of Finland under the Centre of Excellence Programme
2012-2017 (Nuclear and Accelerator Based Physics Programme at JYFL) and under the FIDIPRO
programme, and by the European Union’s Seventh Framework Programme ENSAR (THEXO) under Grant
No. 262010.
We acknowledge the CSC - IT Center for Science Ltd, Finland, for the allocation of computational
resources.

\end{document}